\documentclass{aa}  
 
\usepackage{graphicx}

\usepackage{txfonts}
\usepackage{CJKutf8}
\usepackage{threeparttable}
\usepackage{cases}
\usepackage{verbatim}
\usepackage[colorlinks,citecolor=cyan]{hyperref}

\begin{document}

\title{Radio eclipse of the slowest spinning Galactic-field spider pulsar PSR J1932+2121 and its X-ray emission prospect}

\author{
{Ze-Xin Du \begin{CJK*}{UTF8}{gbsn}(杜泽昕) \end{CJK*}} \inst{1, 2} 
\and {Yun-Wei Yu \begin{CJK*}{UTF8}{gbsn} (俞云伟)\end{CJK*}} \inst{1, 2} \thanks{Corresponding author}
\and {Zong-Lin Yang \begin{CJK*}{UTF8}{gbsn}(杨宗霖) \end{CJK*}} \inst{3, 4}
\and {Aming Chen} \inst{1, 2}
\and {Peng-Fei Wang \begin{CJK*}{UTF8}{gbsn} (王鹏飞) \end{CJK*}} \inst{3, 4, 5}
\and {Jin-Lin Han \begin{CJK*}{UTF8}{gbsn} (韩金林) \end{CJK*}} \inst{3, 4, 5} 
\and {De-Jiang Zhou \begin{CJK*}{UTF8}{gbsn} (周德江) \end{CJK*}} \inst{3, 4, 5}
\and {Xiao-Peng You \begin{CJK*}{UTF8}{gbsn} (游霄鹏) \end{CJK*}} \inst{6}
}

\institute{
    {Institute of Astrophysics, Central China Normal University, Wuhan 430079, China;} \email{yuyw@ccnu.edu.cn}
\and
    {Laboratory for Compact Object Astrophysics and Astronomical Technology, Central China Normal University, Wuhan 430079, China}
\and
    {National Astronomical Observatories, Chinese Academy of Sciences, Jia-20 Datun Road, ChaoYang District, Beijing 100012, China;} \email{hjl@nao.cas.cn}
\and
    {School of Astronomy and Space Science, University of Chinese Academy of Sciences, Beijing 100049, China}   
\and
    {Key Laboratory of Radio Astronomy and Technology, Chinese Academy of Sciences, Beijing 100101, China}
\and
    {School of Physical Science and Technology, Southwest University, Chongqing 400715, China} \email{yxp0910@swu.edu.cn}
}

\date{Received 2 March 2026; accepted 9 June 2026}

\abstract
{
PSR J1932$+$2121 is a newly discovered spider pulsar with a pronounced radio eclipse identified by the Five-hundred-meter Aperture Spherical radio Telescope (FAST); it provides an ideal laboratory for studying the eclipse mechanism and high-energy emission from its intrabinary shock (IBS).
By modeling the orbital-phase-dependent dispersion measure variations and flux profiles during the eclipse region with the wind interaction and IBS geometry, we constrain the system to a nearly edge-on inclination ($i_{\mathrm{o}} \simeq 88.55^{\circ}$) and a weak stellar wind from a low-mass main-sequence companion.
Our analysis of the eclipse mechanism suggests that synchrotron absorption by  nonthermal electrons can reproduce the observed flux variations with reasonable parameters for the eclipsing medium.
We further predict the synchrotron emission from the IBS in PSR J1932$+$2121, showing that its X-ray flux, particularly near the inferior conjunction of the companion star, could be detectable with XMM/EPIC, EP/FXT, or eXTP/SFA and should exhibit double-peaked orbital modulation by Doppler boosting.
These results provide a theoretical framework for understanding this system and for guiding future multiwavelength probes of spider pulsars.
}


\authorrunning{Du et al.}
\titlerunning{Radio eclipse and X-ray emission of PSR J1932+2121}

\maketitle
\nolinenumbers

\section{Introduction}
The Galactic Plane Pulsar Snapshot (GPPS) survey \citep{Han_2021} by the Five-hundred-meter Aperture Spherical radio Telescope (FAST) has discovered a large number of new pulsars; to date,  751 pulsars have been detected \citep{Han_2025}.
Approximately 20\% of these exhibit characteristics consistent with binary systems \citep{WangPF_2024}, thus enabling more detailed studies of binary properties, evolutionary features, and population statistics \citep{Koljonen_2025}. Here, when the binary orbital plane is close to the direction of the line of sight (LOS) and the companion star moves near its inferior conjunction (INFC), the wind material from the companion is expected to obscure the radio emission of the pulsar, producing periodical eclipse features \citep{Kluzniak_1988, Stappers_1996-asps, Guillemot_2019, Nieder_2020}. The study of these eclipses provides valuable insight into the eclipse mechanism and probes the physical conditions of the binary environment \citep{Phinney_1988, Thompson_1994, Polzin_2020, Miao_2023}.

In particular, when the pulsar is a millisecond pulsar (MSP), the companion star can be ablated significantly by pulsar irradiation \citep{Fruchter_1988, Khechinashvili_2000, Chen_2013, Koljonen_2025}, finally becoming a low-mass star in tight, near-circular orbits. Such pulsar binaries are generally called spider pulsar binaries. More specifically, these systems are commonly subdivided into redbacks, with companion masses of $\sim 0.2-0.4 M_{\odot}$, and black widows, whose companions are ultra-low mass, $\sim 0.02-0.05 M_{\odot}$  \citep{Roberts_2013, Polzin_2019}. The strong interaction between the winds of the pulsar and companion can form an intrabinary shock (IBS; e.g., bow shock), which significantly increases the chance of detecting an eclipse phenomenon \citep{Wadiasingh_2017, Du_2023}. Moreover, the resulting cometary structure of the IBS would directly govern the radio eclipse boundary.

In addition to radio eclipses, high-energy emission has also been detected from the interaction of the spider pulsar wind with the evaporating companion material \citep{Ruderman_1989, Huang_2007, Roberts_2014, Karpova_2025, Maksat_2025} as this emission can be naturally interpreted as synchrotron radiation from the IBS zone that is modulated by the orbital motion \citep{Romani_2016, Wadiasingh_2017, Kandel_2019, Martino_2020, Sim_2024}. Therefore, a joint modeling of the radio eclipse and the IBS high-energy emission is essential for constraining the properties of the companion outflow, the wind interaction, and even the dynamical evolution of the pulsar wind.

Among the growing binary samples, a remarkable fraction have been identified as spider pulsars \citep{WangPF_2024}, including PSR J1932$+$2121, which was discovered in the FAST GPPS survey. PSR J1932$+$2121 is distinctive as the slowest spinning Galactic-field spider pulsar currently known, with a spin period of $14.25$ ms, ultra-compact orbit, and pronounced radio eclipses. Therefore, this work was devoted to modeling the radio eclipses of this particular spider pulsar and to predicting its multiwavelength emission properties.
The paper is organized as follows. In Sect. \ref{sect:Data} we describe our observation with FAST and data processing. The flux and dispersion measure (DM) obtained from PSR J1932$+$2121 are presented.
In Sect. \ref{sect:Model} we revisit the eclipse model and constrain the parameters of the binary orbit and the companion star. In Sect. \ref{sect:Xray} we calculate the X-ray emission arising from the IBS with the obtained model parameters, where the dynamical evolution of the pulsar wind is taken into account. Finally, the implications for the eclipse mechanism and the intrabinary environment of PSR J1932$+$2121 are summarized in Sect. \ref{sect:conclusion}.

\section{FAST observations of PSR J1932$+$2121}
\label{sect:Data}

\begin{figure*}[htbp]
\centering
\includegraphics[width=0.45\linewidth]{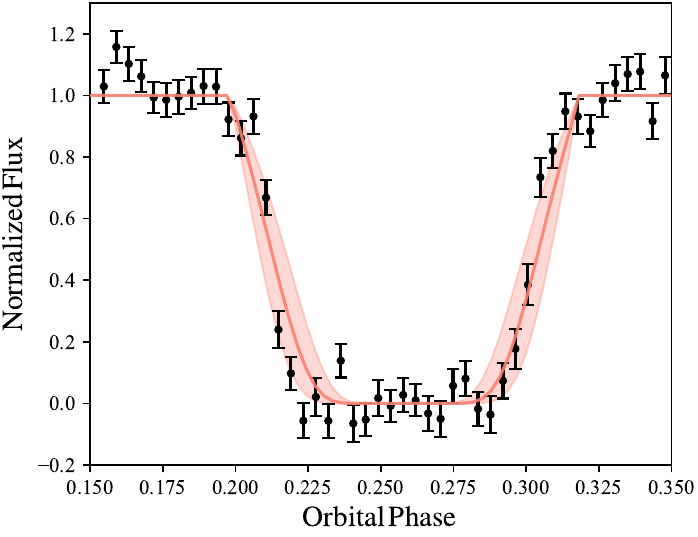}   \includegraphics[width=0.45\linewidth]{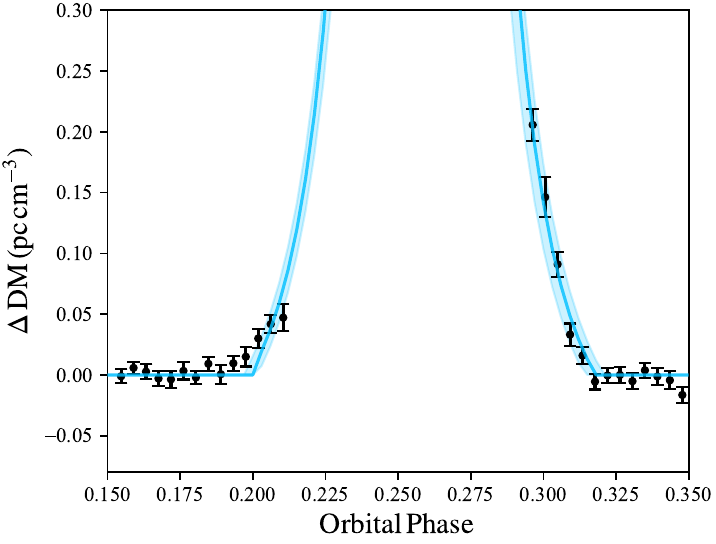}
\caption{Normalized radio light curve of PSR J1932$+$2121 during the eclipse period (\textit{left}) and the corresponding DM variation (\textit{right}). The solid lines give the fitting of the data with the model and 1$\sigma$ confidence regions shown as shaded areas. The adopted parameter values are listed in Table \ref{T1}.}
\label{FluxDM}
\end{figure*}

PSR J1932$+$2121 has a spin period of $14.25$ ms and a spin-down luminosity of 4.8$\times10^{33}$ erg s$^{-1}$ \citep{WangPF_2024} and moves in a 1.94-hour orbit. The minimum mass of the companion is 0.115~$M_\odot$, placing it among spider pulsars \citep{WangPF_2024, Han_2025, Misra_2025}.
However, the nature of the companion is still unclear. Based on the available observations and the phenomenology of spider systems, it is thought to be a low-mass main-sequence star, although its specific type still requires future confirmation \citep{WangPF_2024}.

The radio emission of PSR J1932$+$2121 is regularly eclipsed, most likely by material from its companion near the INFC phase.\footnote{Primary eclipses of pulsar emission are generally expected to occur in the orbital-phase interval near the INFC of the companion, during which the radio flux is strongly suppressed, accompanied by a synchronous DM excess. However, it is still noteworthy that some spider systems could exhibit irregular or weaker eclipse-like features in multiple orbital phases (e.g., PSR B1744-24A, PSR J1717$+$4308A, PSR J1810-1744; \citealt{Bilous_2019, Pan_2020, Kumari_2025}), which might arise from clumps or otherwise complex material in the binary environment. Such complexity has not been discovered in PSR J1932$+$2121.} 
To study these eclipses, we used the two-hour FAST tracking observation on August 11, 2022. By taking the ascending node as phase zero, we determined the INFC of the companion at phase 0.25.  Around this eclipse region, the pulse radio flux changed gradually and an extra time delay $\delta t$ can be seen due to the excess dispersion measure (DM$_{\rm ex}$) as 
\begin{equation}
    \delta t=4.148808(3)\times10^3~{\rm s}\times \left(\frac{{\rm DM}_{\rm ex}}{\rm pc\,cm^{-3}} \right) \left(\frac{f_{\rm obs}}{\rm MHz} \right)^{-2},
\end{equation}
where $f_{\rm obs}$ is the observation frequency 1250 MHz. We extracted the flux density variations using \texttt{psrflux} command in \textsc{Psrchive} package \citep{Hotan+2004PASA...21..302H}, and determined the excess DM using the timing residuals reported by \textsc{Tempo2} \citep{Hobbs+2006MNRAS.369..655H}. The variation in the radio flux of the pulsar is displayed in Fig. \ref{FluxDM}. Accompanying with the suppression of the flux, the DM of the radio emission is increased significantly. 
These features clearly show that the pulsar radio emission is eclipsed by material from the companion outflow.

\begin{table}[htbp]
\centering
\caption{Parameters of PSR J1932$+$2121.}
\label{T1}
\small
\begin{tabular}{lc}
\hline\hline
Parameter & PSR J1932$+$2121 \\
\hline
Orbital period, $P_{\mathrm{orb}}$ (days) & $0.0809^{(a)}$ \\
Spin period, $P_{\mathrm{spin}}$ (ms) & $14.25^{(a)}$ \\
Spin-down luminosity, $L_{\mathrm{sd}}$ ($\mathrm{erg\,s^{-1}}$) & $4.81\times10^{33\,(b)}$ \\
Eccentricity, $e$ ($10^{-5}$) & $2.8^{(a)}$ \\
Binary separation, $a_{\mathrm{orb}}$ (cm) & $6.23\times10^{10\,(c)}$ \\
Distance, $d_{\mathrm{L}}$ (kpc) & $1.5\sim5.1^{(c)}$ \\
\hline
Dispersion measure, DM ($\mathrm{pc\,cm^{-3}}$) & $192.10^{(a)}$ \\
Projected semi-major axis, $x$ (lt-s) & $0.16269^{(a)}$ \\
Mass function, $f_{\mathrm{m}}$ ($10^{-4}\,M_{\odot}$) & $7.063^{(a)}$ \\
Minimum companion mass, $M_{\mathrm{C,min}}$ ($M_{\odot}$) & $0.115^{(a)}$ \\
\hline
\end{tabular}
\tablefoot{The parameters listed in the lower part of the table are provided as basic information for reference and are not used in the present calculations.}
\tablebib{(a)~\citet{WangPF_2024}; (b)~\citet{Koljonen_2025};
(c)~\citet{Misra_2025}.}
\end{table}

\section{The radio eclipse}
\label{sect:Model}

The periodic eclipse of pulsar radio emission can occur due to the orbital motion of the companion star around the pulsar when the LOS direction passes through the companion star and its outflows within an IBS shock \citep{Du_2023}. Such an IBS, also known as a bow shock, can arise from the interaction between the relativistic pulsar wind and the companion wind \citep{Bosch-Ramon_2011, Romani_2016, Wadiasingh_2017, Kandel_2021}.
The geometry of this shock is determined by mechanical pressure balance across the contact discontinuity, and is primarily governed by the momentum flux ratio of the two winds as\begin{equation}
    \eta = \frac{L_{\mathrm{sd}} / c}{\dot{M}_{\mathrm{C}} \, v_{\mathrm{w}}} \label{1},
\end{equation}
where $L_{\mathrm{sd}}$ is the spin-down power of the pulsar, $c$ is the speed of light, $\dot{M}_{\mathrm{C}}$ is the mass-loss rate of the companion star, and $v_{\mathrm{w}}$ is the terminal velocity of this wind. For constants $\dot{M}_{\mathrm{C}}$ and $v_{\mathrm{w}}$, the number density of the companion wind can be written as
\begin{equation}
    n_{\mathrm{w,i}} (r) = n_{\star} \left( \frac{r}{r_{\star}}\right)^{-2} \label{3},
\end{equation}
where the base density $n_{\star}$ at the stellar surface $r_{\mathrm{\star}}$ is given by
$n_{\star} = \dot{M}_{\mathrm{C}} / 4 \pi {r_{\star}}^2 v_{\mathrm{w}} \mu_{\mathrm{i}} m_{\mathrm{p}}$ with $m_{\mathrm{p}}$   the mass of protons and $r$  the distance from the center of the companion. 

A detailed description of the radio eclipse mechanism for spider pulsars was provided by \cite{Thompson_1994}.
It was further ruled out that the eclipse is caused by scattering and refraction \citep{Broderick_2016, Kudale_2020};  
instead, the absorption process could play the most important role in suppressing the pulsar radio emission. Then, with a frequency-dependent absorption coefficient $\alpha(\nu, \, n_{\mathrm{e}})$, we can express the eclipse using the absorption optical depth along the LOS as \citep{Chen_2021}
\begin{equation}
    \tau(\nu) = \int_{l_{\mathrm{p,obs}}}^{\infty} \alpha(\nu, \, n_{\mathrm{e}}) \, \mathrm{d} l \label{tau},
\end{equation}
which strongly depends on the physical process dominating the absorption. 
Here, $l_{\mathrm{p,obs}}$ is aimed at describing the shock cavity size, and $n_{\mathrm{e}}$ is the electron number density, which is determined by the hydrogen abundance of the wind and the distance to the center of the companion star. Specifically, the electron number density can be related to the ion density by $n_{\mathrm{e}} = n_{\mathrm{w, i}}\mu_{\mathrm{i}} / \mu_{\mathrm{e}}$, where $\mu_{\mathrm{i}} \sim 1.29$ and $\mu_e \sim 1.18$ are the typical values for the mean ion molecular weight and electron weight, respectively \citep{Zdziarski_2010}. 
This optical depth determines the attenuation of the pulsar radio emission and is therefore constrained by the observed flux variations during the eclipse.
Meanwhile, the enhanced electron density along the LOS can contribute an extra component of the dispersion measure (DM), which can be expressed as
\begin{equation}
    \Delta \mathrm{DM} = \int_{l_{\mathrm{p,obs}}}^{\infty} n_{\mathrm{e}} \, \mathrm{d} l \label{dm}.
\end{equation}
The observed variations of the radio flux and DM jointly constrain the location and path length of the LOS through the eclipsing medium enveloped by the bow shock, thereby effectively constraining the IBS geometry represented by $l_{\mathrm{p,obs}}$.

For the eclipse mechanism, measurements of magnetic fields in the eclipse medium indicated that the cyclotron--synchrotron absorption mechanism may play a more important role in radio absorption \citep{Polzin_2019, Lin_2023, Wang_2023}. Specifically, the measured magnetic field strength could be too low to meet the requirement of cyclotron absorption \citep{Thompson_1994, LiDz_2019, Kumari_2024}; furthermore, cyclotron absorption cannot account for the observed broadband as it is expected to feature at the cyclotron frequency and its harmonics \citep{Khechinashvili_2000, Kansabanik_2021}. 
Alternatively, some recent studies suggested that the eclipse could be dominated by synchrotron absorption, such as the 4 GHz eclipse of PSR J1908$+$2105 \citep{Ghosh_2025}.
Therefore, in this work we took into account synchrotron absorption with an absorption coefficient given by \citep{YangYP_2016, Ghosh_2025}
\begin{align}
    \alpha_{\mathrm{syn, \, nth}} = & \frac{q_{\mathrm{e}}^{2}}{4 m_{\mathrm{e}} c} 3^{\frac{p+2}{2}} \Gamma \left( \frac{3 p + 2}{12} \right) \Gamma \left( \frac{3 p + 22}{12} \right) \nonumber \\ & \times(\nu_{\mathrm{B}} \,  \mathrm{sin} \,  \theta)^{\frac{p+2}{2}} \nu^{ - \frac{p+4}{2}} f n_{\mathrm{e}}, \label{11}
\end{align}
where $q_{\mathrm{e}}$ and $m_{\mathrm{e}}$ are the electron charge and mass, $\nu_{\mathrm{B}} = q_{\mathrm{e}} B_{\mathrm{m}} / 2 \pi m_{\mathrm{e}} c$ is the Larmor frequency of the electron, $B_{\mathrm{m}}$ is the local magnetic field strength,  $\theta$ is the angle between the LOS and magnetic field lines, $\Gamma$ denotes the Gamma function, and $p$ is the power-law (PL) index of the nonthermal electron distribution. In the above expression, a PL distribution with index $p$ is adopted as 
\begin{align}
    n(\gamma) = \frac{p-1}{\gamma_{\mathrm{min}}^{1-p} - \gamma_{\mathrm{max}}^{1-p}} f_{\mathrm{nth}} n_{\mathrm{e}} \equiv f n_{\mathrm{e}}, \label{10}
\end{align}
for relativistic nonthermal electrons in the medium, where $\gamma_{\mathrm{min}}$ and $\gamma_{\mathrm{max}}$ are the minimum and maximum Lorentz factors, respectively, and $f_{\rm nth}$ represents the fraction of nonthermal electrons. In the following calculations, we took the combination coefficient $f$ as a free parameter and assumed the PL index $p$ to be 2.5 as a fiducial value for the nonthermal electron distribution. In comparison, the contribution of thermal electrons is ignored, which can be safe as long as the electron temperature satisfies $k_{\rm B}T_{\rm e}\lesssim  0.1m_{\rm e}c^2$.

\begin{table}[htbp]
\centering
\caption{Fitting parameters.}
\label{Fit}
\small
\begin{tabular}{lc}
\hline\hline
Parameter & PSR J1932$+$2121 \\
\hline
Inclination angle of observer, $i_{\mathrm{o}}$ ($^\circ$) & $88.55^{+2.12}_{-3.90}$ \\
True anomaly of observer, $\phi_{\mathrm{o}}$ ($^\circ$) & $92.76^{+0.09}_{-0.13}$ \\
Mass-loss rate, $\log_{10}\dot{M}_{\mathrm{C}}$ ($M_{\odot}\,\mathrm{yr^{-1}}$) & $-12.46^{+0.05}_{-0.02}$ \\
Wind velocity, $v_{\mathrm{w}}$ ($10^8\,\mathrm{cm\,s^{-1}}$) & $0.34^{+0.06}_{-0.03}$ \\
Magnetic field perpendicular to LOS, $B_{\mathrm{m}}\sin\theta$ (G) & $2.17^{+0.43}_{-0.26}$ \\
Effective nonthermal fraction, $f$ ($10^{-2}$) & $1.34^{+0.41}_{-0.43}$ \\
\hline
\end{tabular}
\end{table}

In Fig. \ref{FluxDM} we present the fitting results for the normalized flux and DM variations of PSR J1932$+$2121 during the eclipse;   the 1$\sigma$ confidence regions are shown as shaded areas. The corresponding best-fit parameters are summarized in Table \ref{Fit}.
The inferred orbital inclination is approximately ${88.55^{\circ}}^{+2.12^{\circ}}_{-3.90^{\circ}}$, indicating that the orbit of PSR J1932$+$2121 is  viewed nearly edge-on.
The mass-loss rate  and the velocity of the companion wind are constrained to be $10^{-12.46} \,  \mathrm{M_{\odot} \, yr^{-1}}$ and 0.34 $\times$ $10^{8} \, \mathrm{cm \, s^{-1}}$, respectively. These parameter values are consistent with the companion being a low-mass main-sequence star \citep{WangPF_2024}.
Meanwhile, with such a weak stellar wind and a pulsar spin-down luminosity of $4.81 \times 10^{33} \, \mathrm{erg \, s^{-1}}$, the evaporation efficiency is only $\sim 10^{-4}$, thus indicating that ablation of the companion is highly inefficient, which allows the system to remain in a relatively stable state for a long time \citep{Misra_2025}.

\begin{figure*}[htbp]
\centering
\includegraphics[scale = 0.65]{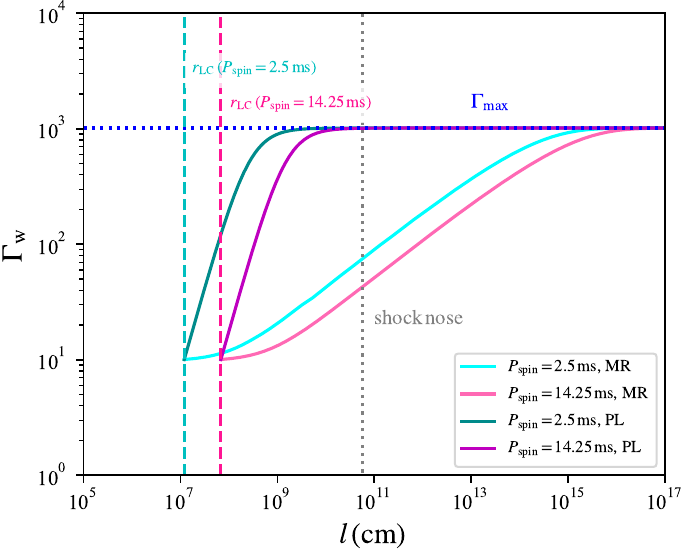}
\includegraphics[scale = 0.65]{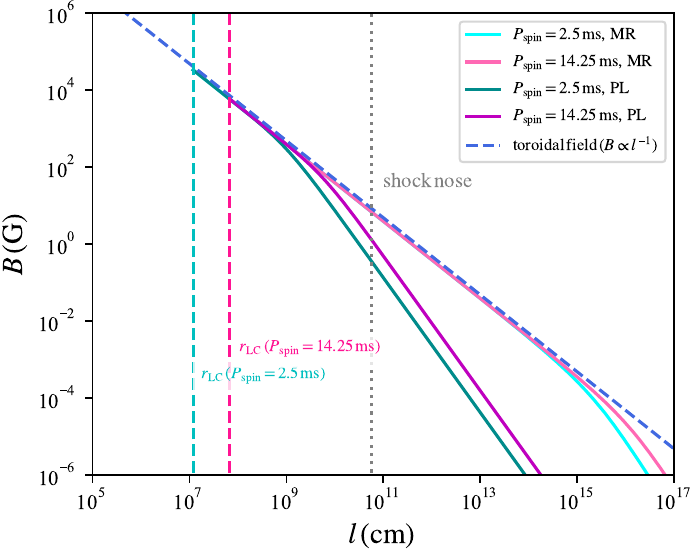}
\caption{Bulk Lorentz factor (\textit{left}) and the magnetic field distribution (\textit{right}) in pulsar wind with distance from the light-cylinder radius $r_{\mathrm{LC}}$ for the MR and PL magnetization decay models with different spin period of $P_{\mathrm{spin}}=2.5$ ms and $14.25$ ms. The dotted line indicates the position of the IBS nose, i.e., the shock stand-off distance ($l_0$). In the left panel, the horizontal dotted line depicts the maximum Lorentz factor $\Gamma_{\max}$, corresponding to complete conversion of the available Poynting flux into bulk kinetic energy. In the right panel, the toroidal field model ($B \propto l^{-1}$) is compared with the MHD-compressed field derived from shock jump conditions.}
\label{mag}
\end{figure*}

\section{Pulsar wind dynamics and X-ray emission prospects}
\label{sect:Xray}

\subsection{Implication for the dynamics of pulsar wind}

Initially, the pulsar wind is dominated by the Poynting flux, and the conversion of electromagnetic energy into bulk kinetic energy during wind expansion reduces magnetization with increasing radius \citep{Bogovalov_1999, Aharonian_2012, Takata_2017, Sullivan_2024}.
The evolution of particles in the pulsar wind can be derived from relativistic magnetohydrodynamics (MHD) by accounting for magnetic reconnection (MR) as the dominant dissipation mechanism \citep{Lyubarsky_2001, Drenkhahn_2002_b, Cortes_2024}. 
In the following, we refer to this case as the MR model.
Under the assumption that the magnetic energy is entirely converted into bulk kinetic energy rather than particle internal energy, the dynamical description can be obtained by numerically solving the equation as follows \citep{Drenkhahn_2002_a}:
\begin{equation}
    \frac{\mathrm{d} \Gamma_{\mathrm{w}}}{\mathrm{d} l} = \frac{2}{c \, \tau} \left[ (1 - \mu^2)(1 + \sigma_{\mathrm{LC}}) \Gamma_{\mathrm{LC}} + \mu^2 \Gamma_{\mathrm{LC}} - \Gamma_{\mathrm{w}} \right] \label{12}.
\end{equation}
Here $\Gamma_{\mathrm{w}}$ is the bulk Lorentz factor of the pulsar wind; $\tau$ is the MR timescale;  $\mu = \mathrm{cos} \, i$ is the magnetic obliquity, with $i$ the angle between the spin and magnetic axes of the pulsar; and
$\sigma_{\mathrm{LC}}$ and $\Gamma_{\mathrm{LC}}$ are the magnetization parameter and the Lorentz factor at light-cylinder radius $r_{\mathrm{LC}}$, respectively. Furthermore, the compression of the magnetic field at the shock can be derived by applying the MHD shock jump conditions with the magnetization in the pulsar wind \citep{Kennel_1984_a, Chen_2019},
\begin{equation}
    B(l) = \sqrt{ \frac{L_{\mathrm{sd}} \sigma}{l^2 c \, (1 + \sigma)} \left(1 + \frac{1}{u^2} \right)} \label{14},
\end{equation}
where $u$ is the radial four velocity.
\cite{Kandel_2019} assumed a toroidal magnetic field structure in the pulsar wind outside the light cylinder as $B(l) = B_0 (l_0 / l)$, where $B_0 = ( 3 L_{\mathrm{sd}} / 2 \, c \, l_0^2)^{1/2}$ is the magnetic field at the nose of the shock $l_0$ \citep{Kandel_2019, Sullivan_2023}.
In the relativistic limit $u \gg 1$ and for a slowly varying $\sigma(l)$, Eq. \ref{14} simplifies to $B \propto l^{-1}$, which exhibits the same radial decrease as the toroidal model.
We found that the two magnetic field descriptions are broadly similar over most of the shock region, but their differences become more pronounced in the distant shock tail, as shown in the right panel of Fig. \ref{mag}.

Meanwhile, some investigations suggest that MR in the striped pulsar wind is not a highly efficient method for converting Poynting energy into bulk kinetic energy, and the magnetization of the pulsar wind is assumed to evolve with radial distance in the form of a PL, which we hereafter refer to as the PL model \citep{Kong_2012, Takata_2017},
\begin{equation}
    \sigma(l) = \sigma_{\mathrm{LC}} \left( \frac{l}{r_{\mathrm{LC}}} \right)^{-\alpha_{\sigma}}  \label{PL},
\end{equation}
where $\alpha_{\sigma}$ is taken to be 1.5.
Correspondingly, according to energy conservation, the Lorentz factor and magnetization in the pulsar wind zone are related by \citep{ChenA_2021}\begin{equation}
\Gamma_{\mathrm{w}}(l) \simeq \Gamma_{\mathrm{LC}}{1 + \sigma_{\mathrm{LC}}\over 1+\sigma(l)}.   \label{Gamma-sigma}
\end{equation}
Moreover, the faster dissipation of magnetic energy in the PL model results in a weaker magnetic field on the shock nose and decays more sharply with distance toward the shock tail.

The resulting Lorentz factor profiles for a spin period of $14.25$ ms (PSR J1932$+$2121) and a typical MSP value of 2.5 ms, corresponding to the median spin period of spider pulsars \citep{Koljonen_2025}, are shown in the left panel of Fig. \ref{mag}.
The PL model indicates that the particles can be fully accelerated at the shock nose regardless of the spin period. In contrast, in MR dissipation, magnetic energy cannot be fully converted into bulk kinetic energy at the shock nose, leading to a lower Lorentz factor for the slower-spinning pulsar.

\subsection{The emission of the IBS}
\begin{figure*}[htbp]
\centering
\includegraphics[scale = 0.65]{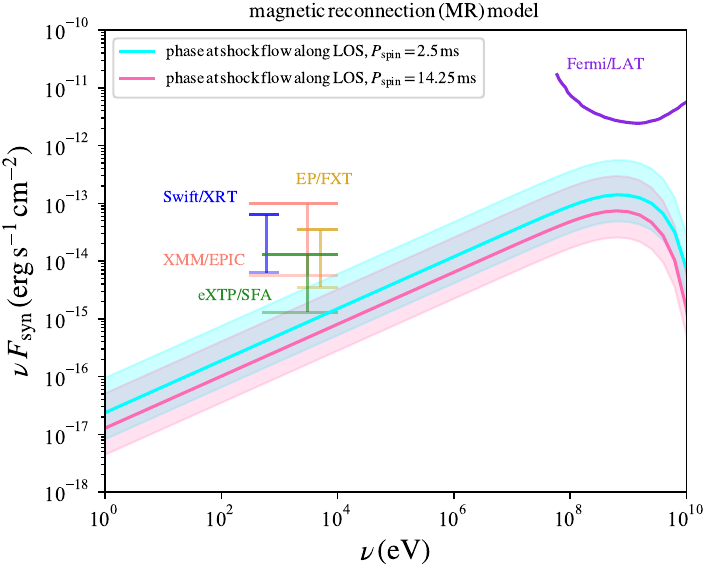}
\includegraphics[scale = 0.65]{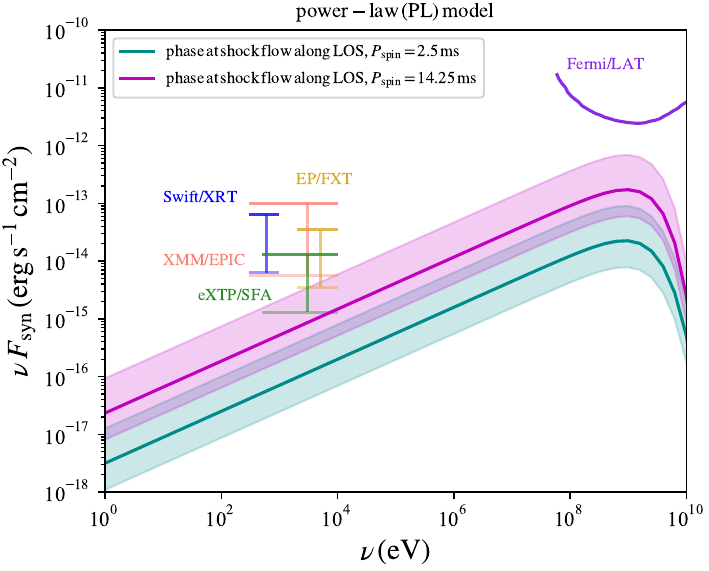}
\caption{SED of PSR J1932$+$2121 at the phase of shock flow along the LOS, shown for the MR (\textit{left}) and PL magnetization decay (\textit{right}) models. The shaded region indicates the uncertainty associated with the source distance, taken to be 1.5 $\sim$ 5.1 kpc by \cite{Misra_2025}, while the solid line running through the middle of each shaded region corresponds to the calculation with the adopted average distance of 3.3 kpc. The sensitivity limits for Swift, EP/FXT, XMM/EPIC, and eXTP/SFA are adopted from \cite{Burrows_2005}, \cite{Yuan_2022}, \cite{Traulsen_2019}, and \cite{ZhangSN_2019}, respectively, assuming representative exposure times of 1--100 ks. The upper limits of \textit{Fermi}/LAT are  from \cite{Werner_2013} and are shown as an additional constraint on the high-energy tail of the spectrum.} 
\label{nu}
\end{figure*}

The pulsar wind is terminated by the stellar outflows, forming a contact discontinuity that separates two shocked regions.
The pulsar wind carries and dissipates most of the Poynting flux energy and is highly magnetized and relativistic downstream of the shock, whereas the companion wind remains comparatively nonrelativistic and weakly magnetized \citep{Wadiasingh_2017, Cortes_2024, Cortes_2025}.
Therefore, we focus on the IBS region where upstream electrons of the pulsar wind are accelerated to ultra-relativistic energies and subsequently cool through adiabatic or radiative processes \citep{Lyubarsky_2003, Sironi_2011, Kandel_2021, Sullivan_2025}.
The accelerated electron--positron pairs are assumed to be distributed as a PL, $ Q(\gamma) = Q_0 \gamma^{-p}$, where $Q_0$ is the injection rate per unit volume and per unit $\gamma$. 
Under the steady-state assumption, the post-shock distribution of shocked electrons can be expressed by the solution of the continuity equation as \citep{Zabalza_2013, Chen_2019}\begin{equation}
    n(\gamma) = \frac{1}{|\dot{\gamma}|} \int_{\gamma} Q(\gamma') \mathrm{d} \gamma', \label{ce}
\end{equation}
where $\dot{\gamma}$ is the total energy loss rate.
By accounting for synchrotron cooling, which dominates the energy losses of electrons traveling downstream in the shock magnetic field \citep{Ghisellini_2013, Cortes_2022, Cortes_2025}, we can calculate the emissivity of the shock by
\begin{equation}
    j(\nu) = \int_{\gamma_{\mathrm{min}}}^{\gamma_{\mathrm{max}}} n(\gamma) P(\gamma) \mathrm{d} \gamma, \label{j_nu}
\end{equation}
where $P(\gamma)$ is the synchrotron power of a single electron.
The minimum Lorentz factor of the shocked electrons is determined by the pulsar wind at the pre-shock region, which is given by
\begin{equation}
\gamma_{\mathrm{min}} \simeq \Gamma_{\mathrm{w}} {p-2 \over p-1},
\end{equation}
which is highly dependent on the bulk Lorentz factor of the injecting pulsar wind. 
The maximum Lorentz factor depends on the acceleration and cooling process, which can be obtained as $\gamma_{\mathrm{max}} = (6 \pi q_{\mathrm{e}} / \sigma_{\mathrm{T}} B)^{1/2}$.
The relativistic bulk motion of the shocked flow causes the emission from the downstream to be strongly beamed, resulting in a Doppler boost at the shock tail when the beaming direction passes through the LOS \citep{Dubus_2010, Wadiasingh_2017, van_2020, Cortes_2025}.
Thus, the total flux from the bow shock can be calculated as \citep{Granot_1999, Kandel_2019}
\begin{equation}
    F(\nu) = \frac{1}{d_{\mathrm{L}}^2} \int_{V} \mathcal D_{\mathrm{}}^2 \, j(\nu/\mathcal D_{\mathrm{}}) \mathrm{d} V, \label{F_nu}
\end{equation}
where $\mathcal D_{\mathrm{}}$ is the Doppler factor determined by the bulk Lorentz factor and the angle between the flow direction and the LOS \citep{Kathirgamaraju_2018, ChenA_2021}.

\subsection{X-ray spectra and light curves}

\begin{figure}[htbp!]
\centering
\includegraphics[scale = 0.62]{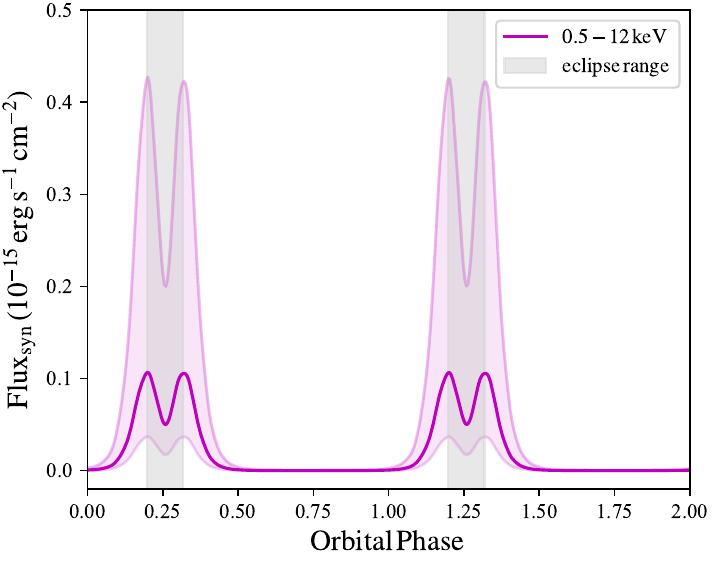}
\caption{Predicted 0.5$-$12 keV X-ray light curve from the IBS region of PSR J1932$+$2121. The gray band represents the orbital phase range of the eclipse, and the purple shaded region indicates the flux uncertainty due to the distance range estimated by \cite{Misra_2025}. The central solid curve corresponds to the calculation with the adopted average distance of 3.3 kpc.}
\label{flux-IBS}
\end{figure}

We further investigated the high-energy synchrotron emission originating from the IBS by the emission model with a magnetization parameter of $10^2$ at $r_{\mathrm{LC}}$ and a particle distribution index of 2.1.
Figure \ref{nu} shows the calculated spectral energy distribution (SED) of PSR J1932$+$2121 at the phase of shock flow along the LOS with strong Doppler boosting \citep{Dubus_2010, Kandel_2019}. 
Although our main interest here is the X-ray detectability of the IBS emission, the modeled SED naturally extends to higher energies, so the \textit{Fermi}/LAT upper limit is also shown as an additional constraint on the high-energy tail of the spectrum.
Since no optical counterpart has been detected, the distance is estimated to be 1.5 $\sim$ 5.1 kpc \citep{Misra_2025}.
Here we used the shaded region to depict the uncertainty of the distance, while the solid curve in Fig. \ref{nu} represents the calculation performed at the average distance of 3.3 kpc.
For PSR J1932$+$2121 with $P_{\mathrm{spin}} = 14.25$ ms, the predicted X-ray SED lies close to or below the current sensitivity limits when the pulsar wind magnetic energy dissipation is governed by the MR model. 
In contrast, if the Poynting flux is dissipated according to the PL model of $\sigma (l)$, the resulting IBS emission becomes significantly brighter and can exceed the typical X-ray SED expected for MSPs.
Therefore, future deep X-ray observations of PSR J1932+2121 will provide a useful test of the pulsar wind Poynting energy dissipation and can place meaningful constraints on the wind dynamics.
Moreover, the integrated X-ray flux in the 0.5$-$12 keV band from the IBS region is shown in Fig. \ref{flux-IBS}.
The double-peaked flux profile over one orbital period arises from Doppler boosting of the shocked flow, which is modeled with a Lorentz factor of 1.8, as the LOS passes across the bow shock twice around the eclipse \citep{Sim_2024}. It is conceivable that when the orbital inclination is small or the half-opening angle of the shock is narrow, the double-peak structure may merge into a single peak \citep{An_2018, Park_2025}.

\section{Summary} \label{sect:conclusion}

The newly discovered redback pulsar binary system with radio eclipse of PSR J1932$+$2121 by FAST provides an opportunity to constrain the binary orbit and the properties of the companion and pulsar. 
Based on fitting the orbital-phase-dependent DM variations by the eclipse model, we inferred a nearly edge-on inclination of $i_{\mathrm{o}} = {88.55^{\circ}} ^{+2.12^{\circ}}_{-3.90^{\circ}}$ and a weak companion stellar wind of $\dot{M}_{\mathrm{C}} \sim 10^{-12.46} \mathrm{M_{\odot} \,  yr^{-1}}$ with $v_w \sim 0.34 \times 10^8 \mathrm{cm \, s}^{-1}$, consistent with a low-mass main-sequence star.

We further modeled the high-energy emission from the IBS zone by explicitly incorporating pulsar wind dynamics, in which MR progressively reduces wind magnetization, and MHD compression at the shock produces a magnetic field scaling similar to the toroidal field assumed by \cite{Kandel_2019}.
The resulting shock synchrotron emission predicts a detectable X-ray flux near the INFC phase with XMM/EPIC, EP/FXT, or eXTP/SFA, while the $\gamma$-ray flux remains below current \textit{Fermi}/LAT limits.
At present, there is no cataloged \textit{Fermi}/LAT source positionally coincident with PSR J1932$+$2121, which is qualitatively consistent with this result.
Taking into account the distance uncertainty, the phase-resolved 0.5$-$12 keV light curve exhibits a characteristic double-peaked light curve around the eclipse range, shaped by Doppler boosting in the post-shock flow.
These high-energy radiation features of PSR J1932$+$2121 can be tested in future observations.
Furthermore, the long spin period of the pulsar ($P_{\rm spin} = 14.25$ ms) and weak companion stellar wind suggest only mild recycling and inefficient late-stage accretion. 
This implies that the actual mass of this neutron star is likely slightly greater than the canonical $1.4 \, \mathrm{M}_\odot$ typically assumed \citep{Ozel_2016}, while still far from the extreme $\gtrsim 2 \, \mathrm{M}_{\odot}$ regime.

\begin{acknowledgements}
We are grateful to Bing-Qing Zhou for helpful discussions and suggestions.
This work is supported by the National Natural Science Foundation of China (grant Nos 12393811 and 12303047), the National SKA Program of China (2020SKA0120300), and the National Key R\&D Program of China (2021YFA0718500).
\end{acknowledgements}

\bibliography{Z-Renference}{}

@ARTICLE{Wadiasingh_2017,
       author = {{Wadiasingh}, Zorawar and {Harding}, Alice K. and {Venter}, Christo and {B{\"o}ttcher}, Markus and {Baring}, Matthew G.},
        title = "{Constraining Relativistic Bow Shock Properties in Rotation-powered Millisecond Pulsar Binaries}",
      journal = {\apj},
         year = 2017,
        month = apr,
       volume = {839},
       number = {2},
          eid = {80},
        pages = {80},
          doi = {10.3847/1538-4357/aa69bf},
archivePrefix = {arXiv},
       eprint = {1703.09560},
 primaryClass = {astro-ph.HE},
       adsurl = {https://ui.adsabs.harvard.edu/abs/2017ApJ...839...80W}
}

@ARTICLE{Hobbs+2006MNRAS.369..655H,
       author = {{Hobbs}, G.~B. and {Edwards}, R.~T. and {Manchester}, R.~N.},
        title = "{TEMPO2, a new pulsar-timing package - I. An overview}",
      journal = {MNRAS},
         year = 2006,
        month = jun,
       volume = {369},
       number = {2},
        pages = {655-672},
          doi = {10.1111/j.1365-2966.2006.10302.x},
archivePrefix = {arXiv},
       eprint = {astro-ph/0603381},
 primaryClass = {astro-ph},
       adsurl = {https://ui.adsabs.harvard.edu/abs/2006MNRAS.369..655H}
}

@ARTICLE{Hotan+2004PASA...21..302H,
       author = {{Hotan}, A.~W. and {van Straten}, W. and {Manchester}, R.~N.},
        title = "{PSRCHIVE and PSRFITS: An Open Approach to Radio Pulsar Data Storage and Analysis}",
      journal = {\pasa},
         year = 2004,
        month = jan,
       volume = {21},
       number = {3},
        pages = {302-309},
          doi = {10.1071/AS04022},
archivePrefix = {arXiv},
       eprint = {astro-ph/0404549},
 primaryClass = {astro-ph},
       adsurl = {https://ui.adsabs.harvard.edu/abs/2004PASA...21..302H}
}

@ARTICLE{Zdziarski_2010,
       author = {{Zdziarski}, Andrzej A. and {Neronov}, Andrii and {Chernyakova}, Maria},
        title = "{A compact pulsar wind nebula model of the {\ensuremath{\gamma}}-ray-loud binary LS I +61{\ensuremath{\circ}}303}",
      journal = {\mnras},
         year = 2010,
        month = apr,
       volume = {403},
       number = {4},
        pages = {1873-1886},
          doi = {10.1111/j.1365-2966.2010.16263.x},
archivePrefix = {arXiv},
       eprint = {0802.1174},
 primaryClass = {astro-ph},
       adsurl = {https://ui.adsabs.harvard.edu/abs/2010MNRAS.403.1873Z}
}

@ARTICLE{Ghosh_2025,
       author = {{Ghosh}, Ankita and {Bhattacharyya}, Bhaswati and {Kumari}, Sangita and {Johnston}, Simon and {Weltevrede}, Patrick and {Roy}, Jayanta},
        title = "{Exploring Unusual High-frequency Eclipses in MSP J1908+2105}",
      journal = {\apj},
         year = 2025,
        month = apr,
       volume = {982},
       number = {2},
          eid = {168},
        pages = {168},
          doi = {10.3847/1538-4357/adb8e0},
archivePrefix = {arXiv},
       eprint = {2502.08357},
 primaryClass = {astro-ph.HE},
       adsurl = {https://ui.adsabs.harvard.edu/abs/2025ApJ...982..168G}
}

@ARTICLE{Kumari_2024,
       author = {{Kumari}, Sangita and {Bhattacharyya}, Bhaswati and {Sharan}, Rahul and {Johnston}, Simon and {Weltevrede}, Patrick and {Stappers}, Benjamin and {Kansabanik}, Devojyoti and {Roy}, Jayanta and {Ghosh}, Ankita},
        title = "{Unveiling Frequency-dependent Eclipsing in Spider Millisecond Pulsars Using Broadband Polarization Observations with the Parkes}",
      journal = {\apj},
         year = 2024,
        month = sep,
       volume = {973},
       number = {1},
          eid = {19},
        pages = {19},
          doi = {10.3847/1538-4357/ad6145},
archivePrefix = {arXiv},
       eprint = {2407.01024},
 primaryClass = {astro-ph.HE},
       adsurl = {https://ui.adsabs.harvard.edu/abs/2024ApJ...973...19K}
}

@ARTICLE{Chen_2021,
       author = {{Chen}, A.~M. and {Guo}, Y.~D. and {Yu}, Y.~W. and {Takata}, J.},
        title = "{Radio absorption in high-mass gamma-ray binaries}",
      journal = {\aap},
         year = 2021,
        month = aug,
       volume = {652},
          eid = {A39},
        pages = {A39},
          doi = {10.1051/0004-6361/202140951},
archivePrefix = {arXiv},
       eprint = {2106.10445},
 primaryClass = {astro-ph.HE},
       adsurl = {https://ui.adsabs.harvard.edu/abs/2021A&A...652A..39C}
}

@ARTICLE{Bosch-Ramon_2011,
       author = {{Bosch-Ramon}, Valenti and {Khangulyan}, Dmitry},
        title = "{Monte Carlo Simulations of Radio Emitting Secondaries in {\ensuremath{\gamma}}-Ray Binaries}",
      journal = {\pasj},
         year = 2011,
        month = oct,
       volume = {63},
        pages = {1023-1033},
          doi = {10.1093/pasj/63.5.1023},
archivePrefix = {arXiv},
       eprint = {1105.2172},
 primaryClass = {astro-ph.HE},
       adsurl = {https://ui.adsabs.harvard.edu/abs/2011PASJ...63.1023B}
}

@ARTICLE{Du_2023,
       author = {{Du}, Ze-Xin and {Yu}, Yun-Wei and {Chen}, A. -Ming and {Wang}, Shuang-Qiang and {Zhou}, Xia and {Zheng}, Xiao-Ping},
        title = "{Constraining the Orbital Inclination and Companion Properties of Three Black Widow Pulsars Detected by FAST}",
      journal = {RAA},
         year = 2023,
        month = dec,
       volume = {23},
       number = {12},
          eid = {125024},
        pages = {125024},
          doi = {10.1088/1674-4527/ad034b},
archivePrefix = {arXiv},
       eprint = {2310.08197},
 primaryClass = {astro-ph.HE},
       adsurl = {https://ui.adsabs.harvard.edu/abs/2023RAA....23l5024D}
}

@BOOK{Ghisellini_2013,
       author = {{Ghisellini}, Gabriele},
        title = "{Radiative Processes in High Energy Astrophysics}",
         year = 2013,
       volume = {873},
          doi = {10.1007/978-3-319-00612-3},
       adsurl = {https://ui.adsabs.harvard.edu/abs/2013LNP...873.....G}
}

@ARTICLE{Sironi_2011,
       author = {{Sironi}, Lorenzo and {Spitkovsky}, Anatoly},
        title = "{Acceleration of Particles at the Termination Shock of a Relativistic Striped Wind}",
      journal = {\apj},
         year = 2011,
        month = nov,
       volume = {741},
       number = {1},
          eid = {39},
        pages = {39},
          doi = {10.1088/0004-637X/741/1/39},
archivePrefix = {arXiv},
       eprint = {1107.0977},
 primaryClass = {astro-ph.HE},
       adsurl = {https://ui.adsabs.harvard.edu/abs/2011ApJ...741...39S}
}

@ARTICLE{Aharonian_2012,
       author = {{Aharonian}, F.~A. and {Bogovalov}, S.~V. and {Khangulyan}, D.},
        title = "{Abrupt acceleration of a `cold' ultrarelativistic wind from the Crab pulsar}",
      journal = {\nat},
         year = 2012,
        month = feb,
       volume = {482},
       number = {7386},
        pages = {507-509},
          doi = {10.1038/nature10793},
       adsurl = {https://ui.adsabs.harvard.edu/abs/2012Natur.482..507A}
}

@ARTICLE{Takata_2017,
       author = {{Takata}, J. and {Cheng}, K.~S.},
        title = "{X-Ray/GeV Emissions from Crab-like Pulsars in the LMC}",
      journal = {\apj},
         year = 2017,
        month = jan,
       volume = {834},
       number = {1},
          eid = {4},
        pages = {4},
          doi = {10.3847/1538-4357/834/1/4},
archivePrefix = {arXiv},
       eprint = {1612.00158},
 primaryClass = {astro-ph.HE},
       adsurl = {https://ui.adsabs.harvard.edu/abs/2017ApJ...834....4T}
}

@ARTICLE{ChenA_2021,
       author = {{Chen}, A.-Ming and {Ng}, Chowing and {Takata}, Jumpei and {Yu}, Yun-Wei},
        title = "{Modeling the high-energy emission from the gamma-ray binary 1FGL J1018.6-5856}",
      journal = {RAA},
         year = 2021,
        month = oct,
       volume = {21},
       number = {8},
          eid = {189},
        pages = {189},
          doi = {10.1088/1674-4527/21/8/189},
archivePrefix = {arXiv},
       eprint = {1703.08080},
 primaryClass = {astro-ph.HE},
       adsurl = {https://ui.adsabs.harvard.edu/abs/2021RAA....21..189C}
}

@ARTICLE{Cortes_2022,
       author = {{Cort{\'e}s}, Jorge and {Sironi}, Lorenzo},
        title = "{Global Kinetic Modeling of the Intrabinary Shock in Spider Pulsars}",
      journal = {\apj},
         year = 2022,
        month = jul,
       volume = {933},
       number = {2},
          eid = {140},
        pages = {140},
          doi = {10.3847/1538-4357/ac74b2},
archivePrefix = {arXiv},
       eprint = {2203.00023},
 primaryClass = {astro-ph.HE},
       adsurl = {https://ui.adsabs.harvard.edu/abs/2022ApJ...933..140C}
}

@ARTICLE{Khechinashvili_2000,
       author = {{Khechinashvili}, David G. and {Melikidze}, George I. and {Gil}, Janusz A.},
        title = "{Nature of Eclipsing Pulsars}",
      journal = {\apj},
         year = 2000,
        month = sep,
       volume = {541},
       number = {1},
        pages = {335-343},
          doi = {10.1086/309408},
archivePrefix = {arXiv},
       eprint = {astro-ph/0001130},
 primaryClass = {astro-ph},
       adsurl = {https://ui.adsabs.harvard.edu/abs/2000ApJ...541..335K}
}

@ARTICLE{Kansabanik_2021,
       author = {{Kansabanik}, Devojyoti and {Bhattacharyya}, Bhaswati and {Roy}, Jayanta and {Stappers}, Benjamin},
        title = "{Unraveling the Eclipse Mechanism of a Binary Millisecond Pulsar Using Broadband Radio Spectra}",
      journal = {\apj},
         year = 2021,
        month = oct,
       volume = {920},
       number = {1},
          eid = {58},
        pages = {58},
          doi = {10.3847/1538-4357/ac19b9},
archivePrefix = {arXiv},
       eprint = {2107.13258},
 primaryClass = {astro-ph.HE},
       adsurl = {https://ui.adsabs.harvard.edu/abs/2021ApJ...920...58K}
}

@ARTICLE{Lin_2023,
       author = {{Lin}, F.~X. and {Main}, R.~A. and {Jow}, D. and {Li}, D.~Z. and {Pen}, U. -L. and {van Kerkwijk}, M.~H.},
        title = "{Plasma lensing near the eclipses of the Black Widow pulsar B1957+20}",
      journal = {\mnras},
         year = 2023,
        month = feb,
       volume = {519},
       number = {1},
        pages = {121-135},
          doi = {10.1093/mnras/stac3456},
archivePrefix = {arXiv},
       eprint = {2208.13868},
 primaryClass = {astro-ph.HE},
       adsurl = {https://ui.adsabs.harvard.edu/abs/2023MNRAS.519..121L}
}

@ARTICLE{LiDz_2019,
       author = {{Li}, Dongzi and {Lin}, Fang Xi and {Main}, Robert and {Pen}, Ue-Li and {van Kerkwijk}, Marten H. and {Yang}, I. -Sheng},
        title = "{Constraining magnetic fields through plasma lensing: application to the Black Widow pulsar}",
      journal = {\mnras},
         year = 2019,
        month = apr,
       volume = {484},
       number = {4},
        pages = {5723-5733},
          doi = {10.1093/mnras/stz374},
archivePrefix = {arXiv},
       eprint = {1809.10812},
 primaryClass = {astro-ph.HE},
       adsurl = {https://ui.adsabs.harvard.edu/abs/2019MNRAS.484.5723L}
}

@ARTICLE{Thompson_1994,
       author = {{Thompson}, C. and {Blandford}, R.~D. and {Evans}, Charles R. and {Phinney}, E.~S.},
        title = "{Physical Processes in Eclipsing Pulsars: Eclipse Mechanisms and Diagnostics}",
      journal = {\apj},
         year = 1994,
        month = feb,
       volume = {422},
        pages = {304},
          doi = {10.1086/173728},
       adsurl = {https://ui.adsabs.harvard.edu/abs/1994ApJ...422..304T}
}

@ARTICLE{Romani_2016,
       author = {{Romani}, Roger W. and {Sanchez}, Nicolas},
        title = "{Intra-binary Shock Heating of Black Widow Companions}",
      journal = {\apj},
         year = 2016,
        month = sep,
       volume = {828},
       number = {1},
          eid = {7},
        pages = {7},
          doi = {10.3847/0004-637X/828/1/7},
archivePrefix = {arXiv},
       eprint = {1606.03518},
 primaryClass = {astro-ph.HE},
       adsurl = {https://ui.adsabs.harvard.edu/abs/2016ApJ...828....7R}
}

@ARTICLE{Kandel_2021,
       author = {{Kandel}, D. and {Romani}, Roger W. and {An}, Hongjun},
        title = "{XMM-Newton Observes the Intrabinary Shock of PSR J1959+2048}",
      journal = {\apjl},
         year = 2021,
        month = aug,
       volume = {917},
       number = {1},
          eid = {L13},
        pages = {L13},
          doi = {10.3847/2041-8213/ac15f7},
archivePrefix = {arXiv},
       eprint = {2107.08156},
 primaryClass = {astro-ph.HE},
       adsurl = {https://ui.adsabs.harvard.edu/abs/2021ApJ...917L..13K}
}

@ARTICLE{Zabalza_2013,
       author = {{Zabalza}, V. and {Bosch-Ramon}, V. and {Aharonian}, F. and {Khangulyan}, D.},
        title = "{Unraveling the high-energy emission components of gamma-ray binaries}",
      journal = {\aap},
         year = 2013,
        month = mar,
       volume = {551},
          eid = {A17},
        pages = {A17},
          doi = {10.1051/0004-6361/201220589},
archivePrefix = {arXiv},
       eprint = {1212.3222},
 primaryClass = {astro-ph.HE},
       adsurl = {https://ui.adsabs.harvard.edu/abs/2013A&A...551A..17Z}
}

@ARTICLE{Chen_2019,
       author = {{Chen}, A.~M. and {Takata}, J. and {Yi}, S.~X. and {Yu}, Y.~W. and {Cheng}, K.~S.},
        title = "{Modelling multiwavelength emissions from PSR B1259-63/LS 2883: Effects of the stellar disc on shock radiations}",
      journal = {\aap},
         year = 2019,
        month = jul,
       volume = {627},
          eid = {A87},
        pages = {A87},
          doi = {10.1051/0004-6361/201935166},
archivePrefix = {arXiv},
       eprint = {1904.07527},
 primaryClass = {astro-ph.HE},
       adsurl = {https://ui.adsabs.harvard.edu/abs/2019A&A...627A..87C}
}

@ARTICLE{Wang_2023,
       author = {{Wang}, S.~Q. and {Wang}, J.~B. and {Li}, D.~Z. and {Yao}, J.~M. and {Manchester}, R.~N. and {Hobbs}, G. and {Wang}, N. and {Dai}, S. and {Xu}, H. and {Luo}, R. and {Feng}, Y. and {Wang}, W.~Y. and {Li}, D. and {Yu}, Y.~W. and {Du}, Z.~X. and {Niu}, C.~H. and {Zhang}, S.~B. and {Zhang}, C.~M.},
        title = "{Change of Rotation Measure during the Eclipse of a Black Widow PSR J2051-0827}",
      journal = {\apj},
         year = 2023,
        month = sep,
       volume = {955},
       number = {1},
          eid = {36},
        pages = {36},
          doi = {10.3847/1538-4357/acea81},
archivePrefix = {arXiv},
       eprint = {2307.13198},
 primaryClass = {astro-ph.HE},
       adsurl = {https://ui.adsabs.harvard.edu/abs/2023ApJ...955...36W}
}

@ARTICLE{Polzin_2019,
       author = {{Polzin}, E.~J. and {Breton}, R.~P. and {Stappers}, B.~W. and {Bhattacharyya}, B. and {Janssen}, G.~H. and {Os{\l}owski}, S. and {Roberts}, M.~S.~E. and {Sobey}, C.},
        title = "{Long-term variability of a black widow's eclipses - A decade of PSR J2051-0827}",
      journal = {\mnras},
         year = 2019,
        month = nov,
       volume = {490},
       number = {1},
        pages = {889-908},
          doi = {10.1093/mnras/stz2579},
archivePrefix = {arXiv},
       eprint = {1909.06130},
 primaryClass = {astro-ph.HE},
       adsurl = {https://ui.adsabs.harvard.edu/abs/2019MNRAS.490..889P}
}

@ARTICLE{Kong_2012,
       author = {{Kong}, S.~W. and {Cheng}, K.~S. and {Huang}, Y.~F.},
        title = "{Modeling the Multiwavelength Light Curves of PSR B1259-63/LS 2883. II. The Effects of Anisotropic Pulsar Wind and Doppler Boosting}",
      journal = {\apj},
         year = 2012,
        month = jul,
       volume = {753},
       number = {2},
          eid = {127},
        pages = {127},
          doi = {10.1088/0004-637X/753/2/127},
archivePrefix = {arXiv},
       eprint = {1205.2147},
 primaryClass = {astro-ph.HE},
       adsurl = {https://ui.adsabs.harvard.edu/abs/2012ApJ...753..127K}
}

@ARTICLE{Drenkhahn_2002_b,
       author = {{Drenkhahn}, G. and {Spruit}, H.~C.},
        title = "{Efficient acceleration and radiation in Poynting flux powered GRB outflows}",
      journal = {\aap},
         year = 2002,
        month = sep,
       volume = {391},
        pages = {1141-1153},
          doi = {10.1051/0004-6361:20020839},
archivePrefix = {arXiv},
       eprint = {astro-ph/0202387},
 primaryClass = {astro-ph},
       adsurl = {https://ui.adsabs.harvard.edu/abs/2002A&A...391.1141D}
}

@ARTICLE{Drenkhahn_2002_a,
       author = {{Drenkhahn}, G.},
        title = "{Acceleration of GRB outflows by Poynting flux dissipation}",
      journal = {\aap},
         year = 2002,
        month = may,
       volume = {387},
        pages = {714-724},
          doi = {10.1051/0004-6361:20020390},
archivePrefix = {arXiv},
       eprint = {astro-ph/0112509},
 primaryClass = {astro-ph},
       adsurl = {https://ui.adsabs.harvard.edu/abs/2002A&A...387..714D}
}

@ARTICLE{Lyubarsky_2001,
       author = {{Lyubarsky}, Y. and {Kirk}, J.~G.},
        title = "{Reconnection in a Striped Pulsar Wind}",
      journal = {\apj},
         year = 2001,
        month = jan,
       volume = {547},
       number = {1},
        pages = {437-448},
          doi = {10.1086/318354},
archivePrefix = {arXiv},
       eprint = {astro-ph/0009270},
 primaryClass = {astro-ph},
       adsurl = {https://ui.adsabs.harvard.edu/abs/2001ApJ...547..437L}
}

@ARTICLE{Kennel_1984_a,
       author = {{Kennel}, C.~F. and {Coroniti}, F.~V.},
        title = "{Magnetohydrodynamic model of Crab nebula radiation.}",
      journal = {\apj},
         year = 1984,
        month = aug,
       volume = {283},
        pages = {710-730},
          doi = {10.1086/162357},
       adsurl = {https://ui.adsabs.harvard.edu/abs/1984ApJ...283..710K}
}

@ARTICLE{Sullivan_2024,
       author = {{Sullivan}, Andrew G. and {Romani}, Roger W.},
        title = "{The Intrabinary Shock and Companion Star of Redback Pulsar J2215+5135}",
      journal = {\apj},
         year = 2024,
        month = oct,
       volume = {974},
       number = {2},
          eid = {315},
        pages = {315},
          doi = {10.3847/1538-4357/ad4d85},
archivePrefix = {arXiv},
       eprint = {2405.13889},
 primaryClass = {astro-ph.HE},
       adsurl = {https://ui.adsabs.harvard.edu/abs/2024ApJ...974..315S}
}

@ARTICLE{Cortes_2024,
       author = {{Cort{\'e}s}, Jorge and {Sironi}, Lorenzo},
        title = "{Particle acceleration and non-thermal emission at the intrabinary shock of spider pulsars - I. Non-radiative simulations}",
      journal = {\mnras},
         year = 2024,
        month = nov,
       volume = {534},
       number = {3},
        pages = {2551-2565},
          doi = {10.1093/mnras/stae2278},
archivePrefix = {arXiv},
       eprint = {2404.03700},
 primaryClass = {astro-ph.HE},
       adsurl = {https://ui.adsabs.harvard.edu/abs/2024MNRAS.534.2551C}
}

@ARTICLE{Cortes_2025,
       author = {{Cort{\'e}s}, Jorge and {Sironi}, Lorenzo},
        title = "{Particle acceleration and non-thermal emission at the intrabinary shock of spider pulsars {\textendash} II. Fast-cooling simulations}",
      journal = {\mnras},
         year = 2025,
        month = sep,
       volume = {542},
       number = {2},
        pages = {917-926},
          doi = {10.1093/mnras/staf284},
archivePrefix = {arXiv},
       eprint = {2501.08387},
 primaryClass = {astro-ph.HE},
       adsurl = {https://ui.adsabs.harvard.edu/abs/2025MNRAS.542..917C}
}

@ARTICLE{Sullivan_2025,
       author = {{Sullivan}, Andrew G. and {Romani}, Roger W.},
        title = "{High-energy Emission from the Intrabinary Shocks in Redback Pulsars}",
      journal = {\apj},
         year = 2025,
        month = may,
       volume = {984},
       number = {2},
          eid = {146},
        pages = {146},
          doi = {10.3847/1538-4357/adc720},
archivePrefix = {arXiv},
       eprint = {2503.24384},
 primaryClass = {astro-ph.HE},
       adsurl = {https://ui.adsabs.harvard.edu/abs/2025ApJ...984..146S}
}

@ARTICLE{Sullivan_2023,
       author = {{Sullivan}, Andrew G. and {Romani}, Roger W.},
        title = "{Polarization of Intrabinary Shock Emission in Spider Pulsars}",
      journal = {\apj},
         year = 2023,
        month = dec,
       volume = {959},
       number = {2},
          eid = {81},
        pages = {81},
          doi = {10.3847/1538-4357/ad09ae},
archivePrefix = {arXiv},
       eprint = {2311.03464},
 primaryClass = {astro-ph.HE},
       adsurl = {https://ui.adsabs.harvard.edu/abs/2023ApJ...959...81S}
}

@ARTICLE{Lyubarsky_2003,
       author = {{Lyubarsky}, Y.~E.},
        title = "{The termination shock in a striped pulsar wind}",
      journal = {\mnras},
         year = 2003,
        month = oct,
       volume = {345},
       number = {1},
        pages = {153-160},
          doi = {10.1046/j.1365-8711.2003.06927.x},
archivePrefix = {arXiv},
       eprint = {astro-ph/0306435},
 primaryClass = {astro-ph},
       adsurl = {https://ui.adsabs.harvard.edu/abs/2003MNRAS.345..153L}
}

@ARTICLE{Sim_2024,
       author = {{Sim}, Minju and {An}, Hongjun and {Wadiasingh}, Zorawar},
        title = "{Modeling X-Ray and Gamma-Ray Emission from Redback Pulsar Binaries}",
      journal = {\apj},
         year = 2024,
        month = apr,
       volume = {964},
       number = {2},
          eid = {109},
        pages = {109},
          doi = {10.3847/1538-4357/ad25fb},
archivePrefix = {arXiv},
       eprint = {2402.02674},
 primaryClass = {astro-ph.HE},
       adsurl = {https://ui.adsabs.harvard.edu/abs/2024ApJ...964..109S}
}

@ARTICLE{Bogovalov_1999,
       author = {{Bogovalov}, S.~V.},
        title = "{On the physics of cold MHD winds from oblique rotators}",
      journal = {\aap},
         year = 1999,
        month = sep,
       volume = {349},
        pages = {1017-1026},
archivePrefix = {arXiv},
       eprint = {astro-ph/9907051},
 primaryClass = {astro-ph},
       adsurl = {https://ui.adsabs.harvard.edu/abs/1999A&A...349.1017B}
}

@ARTICLE{Kandel_2019,
       author = {{Kandel}, D. and {Romani}, Roger W. and {An}, Hongjun},
        title = "{The Synchrotron Emission Pattern of Intrabinary Shocks}",
      journal = {\apj},
         year = 2019,
        month = jul,
       volume = {879},
       number = {2},
          eid = {73},
        pages = {73},
          doi = {10.3847/1538-4357/ab24d9},
archivePrefix = {arXiv},
       eprint = {1905.12591},
 primaryClass = {astro-ph.HE},
       adsurl = {https://ui.adsabs.harvard.edu/abs/2019ApJ...879...73K}
}

@ARTICLE{Park_2025,
       author = {{Park}, Jaegeun and {Kim}, Chanho and {An}, Hongjun and {Wadiasingh}, Zorawar},
        title = "{Revisiting the Intrabinary Shock Model for Millisecond Pulsar Binaries: Radiative Losses and Long-Term Variability}",
      journal = {Astronomische Nachrichten},
         year = 2025,
        month = jan,
       volume = {346},
       number = {1},
          eid = {e20240099},
        pages = {e20240099},
          doi = {10.1002/asna.20240099},
archivePrefix = {arXiv},
       eprint = {2411.05290},
 primaryClass = {astro-ph.HE},
       adsurl = {https://ui.adsabs.harvard.edu/abs/2025AN....34640099P}
}

@ARTICLE{Dubus_2010,
       author = {{Dubus}, G. and {Cerutti}, B. and {Henri}, G.},
        title = "{Relativistic Doppler-boosted emission in gamma-ray binaries}",
      journal = {\aap},
         year = 2010,
        month = jun,
       volume = {516},
          eid = {A18},
        pages = {A18},
          doi = {10.1051/0004-6361/201014023},
archivePrefix = {arXiv},
       eprint = {1004.0511},
 primaryClass = {astro-ph.HE},
       adsurl = {https://ui.adsabs.harvard.edu/abs/2010A&A...516A..18D}
}

@ARTICLE{van_2020,
       author = {{van der Merwe}, C.~J.~T. and {Wadiasingh}, Z. and {Venter}, C. and {Harding}, A.~K. and {Baring}, M.~G.},
        title = "{X-Ray through Very High Energy Intrabinary Shock Emission from Black Widows and Redbacks}",
      journal = {\apj},
         year = 2020,
        month = dec,
       volume = {904},
       number = {2},
          eid = {91},
        pages = {91},
          doi = {10.3847/1538-4357/abbdfb},
archivePrefix = {arXiv},
       eprint = {2010.01125},
 primaryClass = {astro-ph.HE},
       adsurl = {https://ui.adsabs.harvard.edu/abs/2020ApJ...904...91V}
}

@ARTICLE{Granot_1999,
       author = {{Granot}, Jonathan and {Piran}, Tsvi and {Sari}, Re'em},
        title = "{Images and Spectra from the Interior of a Relativistic Fireball}",
      journal = {\apj},
         year = 1999,
        month = mar,
       volume = {513},
       number = {2},
        pages = {679-689},
          doi = {10.1086/306884},
archivePrefix = {arXiv},
       eprint = {astro-ph/9806192},
 primaryClass = {astro-ph},
       adsurl = {https://ui.adsabs.harvard.edu/abs/1999ApJ...513..679G}
}

@ARTICLE{Kathirgamaraju_2018,
       author = {{Kathirgamaraju}, Adithan and {Barniol Duran}, Rodolfo and {Giannios}, Dimitrios},
        title = "{Off-axis short GRBs from structured jets as counterparts to GW events}",
      journal = {\mnras},
         year = 2018,
        month = jan,
       volume = {473},
       number = {1},
        pages = {L121-L125},
          doi = {10.1093/mnrasl/slx175},
archivePrefix = {arXiv},
       eprint = {1708.07488},
 primaryClass = {astro-ph.HE},
       adsurl = {https://ui.adsabs.harvard.edu/abs/2018MNRAS.473L.121K}
}

@ARTICLE{Misra_2025,
       author = {{Misra}, Devina and {Koljonen}, Karri I.~I. and {Linares}, Manuel},
        title = "{The slowest spinning Galactic-field spider PSR J1932+2121: a history of inefficient mass transfer}",
      journal = {\mnras},
         year = 2025,
        month = jul,
       volume = {541},
       number = {1},
        pages = {L58-L64},
          doi = {10.1093/mnrasl/slaf054},
archivePrefix = {arXiv},
       eprint = {2504.05372},
 primaryClass = {astro-ph.HE},
       adsurl = {https://ui.adsabs.harvard.edu/abs/2025MNRAS.541L..58M}
}

@ARTICLE{WangPF_2024,
       author = {{Wang}, P.~F. and {Han}, J.~L. and {Yang}, Z.~L. and {Wang}, T. and {Wang}, C. and {Su}, W.~Q. and {Xu}, J. and {Zhou}, D.~J. and {Yan}, Yi and {Jing}, W.~C. and {Cai}, N.~N. and {Yuan}, J.~P. and {Xu}, R.~X. and {Wang}, H.~G. and {You}, X.~P.},
        title = "{The FAST Galactic Plane Pulsar Snapshot Survey. VIII. 116 Binary Pulsars}",
      journal = {RAA},
         year = 2025,
        month = jan,
       volume = {25},
       number = {1},
          eid = {014003},
        pages = {014003},
          doi = {10.1088/1674-4527/ada3b8},
archivePrefix = {arXiv},
       eprint = {2412.03062},
 primaryClass = {astro-ph.HE},
       adsurl = {https://ui.adsabs.harvard.edu/abs/2025RAA....25a4003W}
}

@ARTICLE{Koljonen_2025,
       author = {{Koljonen}, Karri I.~I. and {Linares}, Manuel},
        title = "{SpiderCat: A Catalog of Compact Binary Millisecond Pulsars}",
      journal = {\apj},
         year = 2025,
        month = nov,
       volume = {994},
       number = {1},
          eid = {8},
        pages = {8},
          doi = {10.3847/1538-4357/ae08a5},
archivePrefix = {arXiv},
       eprint = {2505.11691},
 primaryClass = {astro-ph.HE},
       adsurl = {https://ui.adsabs.harvard.edu/abs/2025ApJ...994....8K}
}

@ARTICLE{An_2018,
       author = {{An}, Hongjun and {Romani}, Roger W. and {Kerr}, Matthew},
        title = "{Signatures of Intra-binary Shock Emission in the Black Widow Pulsar Binary PSR J2241-5236}",
      journal = {\apjl},
         year = 2018,
        month = nov,
       volume = {868},
       number = {1},
          eid = {L8},
        pages = {L8},
          doi = {10.3847/2041-8213/aaedaf},
archivePrefix = {arXiv},
       eprint = {1810.13055},
 primaryClass = {astro-ph.HE},
       adsurl = {https://ui.adsabs.harvard.edu/abs/2018ApJ...868L...8A}
}

@INCOLLECTION{Yuan_2022,
       author = {{Yuan}, Weimin and {Zhang}, Chen and {Chen}, Yong and {Ling}, Zhixing},
        title = "{The Einstein Probe Mission}",
    booktitle = {Handbook of X-ray and Gamma-ray Astrophysics},
         year = 2022,
       editor = {{Bambi}, Cosimo and {Sangangelo}, Andrea},
          eid = {86},
        pages = {86},
          doi = {10.1007/978-981-16-4544-0_151-1},
       adsurl = {https://ui.adsabs.harvard.edu/abs/2022hxga.book...86Y}
}

@ARTICLE{Traulsen_2019,
       author = {{Traulsen}, I. and {Schwope}, A.~D. and {Lamer}, G. and {Ballet}, J. and {Carrera}, F. and {Coriat}, M. and {Freyberg}, M.~J. and {Michel}, L. and {Motch}, C. and {Rosen}, S.~R. and {Webb}, N. and {Ceballos}, M.~T. and {Koliopanos}, F. and {Kurpas}, J. and {Page}, M.~J. and {Watson}, M.~G.},
        title = "{The XMM-Newton serendipitous survey. VIII. The first XMM-Newton serendipitous source catalogue from overlapping observations}",
      journal = {\aap},
         year = 2019,
        month = apr,
       volume = {624},
          eid = {A77},
        pages = {A77},
          doi = {10.1051/0004-6361/201833938},
archivePrefix = {arXiv},
       eprint = {1807.09178},
 primaryClass = {astro-ph.HE},
       adsurl = {https://ui.adsabs.harvard.edu/abs/2019A&A...624A..77T}
}

@ARTICLE{Werner_2013,
       author = {{Werner}, M. and {Reimer}, O. and {Reimer}, A. and {Egberts}, K.},
        title = "{Fermi-LAT upper limits on gamma-ray emission from colliding wind binaries}",
      journal = {\aap},
         year = 2013,
        month = jul,
       volume = {555},
          eid = {A102},
        pages = {A102},
          doi = {10.1051/0004-6361/201220502},
archivePrefix = {arXiv},
       eprint = {1308.2573},
 primaryClass = {astro-ph.HE},
       adsurl = {https://ui.adsabs.harvard.edu/abs/2013A&A...555A.102W}
}

@INPROCEEDINGS{Roberts_2013,
       author = {{Roberts}, Mallory S.~E.},
        title = "{Surrounded by spiders! New black widows and redbacks in the Galactic field}",
    booktitle = {Neutron Stars and Pulsars: Challenges and Opportunities after 80 years},
         year = 2013,
       editor = {{van Leeuwen}, Joeri},
       volume = {291},
        month = mar,
        pages = {127-132},
          doi = {10.1017/S174392131202337X},
archivePrefix = {arXiv},
       eprint = {1210.6903},
 primaryClass = {astro-ph.HE},
       adsurl = {https://ui.adsabs.harvard.edu/abs/2013IAUS..291..127R}
}

@ARTICLE{Chen_2013,
       author = {{Chen}, Hai-Liang and {Chen}, Xuefei and {Tauris}, Thomas M. and {Han}, Zhanwen},
        title = "{Formation of Black Widows and Redbacks{\textemdash}Two Distinct Populations of Eclipsing Binary Millisecond Pulsars}",
      journal = {\apj},
         year = 2013,
        month = sep,
       volume = {775},
       number = {1},
          eid = {27},
        pages = {27},
          doi = {10.1088/0004-637X/775/1/27},
archivePrefix = {arXiv},
       eprint = {1308.4107},
 primaryClass = {astro-ph.SR},
       adsurl = {https://ui.adsabs.harvard.edu/abs/2013ApJ...775...27C}
}

@ARTICLE{Fruchter_1988,
       author = {{Fruchter}, A.~S. and {Gunn}, J.~E. and {Lauer}, T.~R. and {Dressler}, A.},
        title = "{Optical detection and characterization of the eclipsing pulsar's companion}",
      journal = {\nat},
         year = 1988,
        month = aug,
       volume = {334},
       number = {6184},
        pages = {686-689},
          doi = {10.1038/334686a0},
       adsurl = {https://ui.adsabs.harvard.edu/abs/1988Natur.334..686F}
}

@ARTICLE{Ruderman_1989,
       author = {{Ruderman}, M. and {Shaham}, J. and {Tavani}, M.},
        title = "{Accretion Turnoff and Rapid Evaporation of Very Light Secondaries in Low-Mass X-Ray Binaries}",
      journal = {\apj},
         year = 1989,
        month = jan,
       volume = {336},
        pages = {507},
          doi = {10.1086/167029},
       adsurl = {https://ui.adsabs.harvard.edu/abs/1989ApJ...336..507R}
}

@ARTICLE{Karpova_2025,
       author = {{Karpova}, A.~V. and {Zharikov}, S.~V. and {Zyuzin}, D.~A. and {Kirichenko}, A. Yu. and {Shibanov}, Yu. A. and {M{\'a}rquez}, I.~F.},
        title = "{4FGL J1544.2‑2554: A new spider pulsar candidate}",
      journal = {\aap},
         year = 2025,
        month = jan,
       volume = {693},
          eid = {A158},
        pages = {A158},
          doi = {10.1051/0004-6361/202452333},
archivePrefix = {arXiv},
       eprint = {2411.16350},
 primaryClass = {astro-ph.HE},
       adsurl = {https://ui.adsabs.harvard.edu/abs/2025A&A...693A.158K}
}

@ARTICLE{Huang_2007,
       author = {{Huang}, H.~H. and {Becker}, W.},
        title = "{XMM-Newton observations of PSR B1957+20}",
      journal = {\aap},
         year = 2007,
        month = feb,
       volume = {463},
       number = {2},
        pages = {L5-L8},
          doi = {10.1051/0004-6361:20066568},
archivePrefix = {arXiv},
       eprint = {astro-ph/0612594},
 primaryClass = {astro-ph},
       adsurl = {https://ui.adsabs.harvard.edu/abs/2007A&A...463L...5H}
}

@ARTICLE{Roberts_2014,
       author = {{Roberts}, M.~S.~E. and {Mclaughlin}, M.~A. and {Gentile}, P. and {Aliu}, E. and {Hessels}, J.~W.~T. and {Ransom}, S.~M. and {Ray}, P.~S.},
        title = "{Intrabinary shock emission from ``black widows`` and ``redbacks''}",
      journal = {Astronomische Nachrichten},
         year = 2014,
        month = mar,
       volume = {335},
       number = {3},
        pages = {313-317},
          doi = {10.1002/asna.201312038},
archivePrefix = {arXiv},
       eprint = {1402.5507},
 primaryClass = {astro-ph.HE},
       adsurl = {https://ui.adsabs.harvard.edu/abs/2014AN....335..313R}
}

@ARTICLE{Kluzniak_1988,
       author = {{Kluzniak}, W. and {Ruderman}, M. and {Shaham}, J. and {Tavani}, M.},
        title = "{Nature and evolution of the eclipsing millisecond binary pulsar PSR1957 + 20}",
      journal = {\nat},
         year = 1988,
        month = jul,
       volume = {334},
       number = {6179},
        pages = {225-227},
          doi = {10.1038/334225a0},
       adsurl = {https://ui.adsabs.harvard.edu/abs/1988Natur.334..225K}
}
\bibliographystyle{aa}

\end{document}